# Title: MOONSHOTS FOR AGING

**Authors:** Sandeep Kumar[1], Timothy R. Peterson[1]*

**Affiliation:** [1] Department of Internal Medicine, Division of Bone & Mineral Diseases, Department of Genetics, Institute for Public Health, Washington University School of Medicine, BJC Institute of Health, 425 S. Euclid Ave., St. Louis, MO 63110, USA. * Correspondence to: timrpeterson@wustl.edu (T.R.P).

**Abstract**
As the global population ages, there is increased interest in living longer and improving one's quality of life in later years. However, studying aging – the decline in body function – is expensive and time-consuming. And despite research success to make model organisms live longer, there still aren't really any feasible solutions for delaying aging in humans. With space travel, scientists couldn't know what it would take to get to the moon. They had to extrapolate from theory and shorter-range tests. Perhaps with aging, we need a similar moonshot philosophy. And though "shot" might imply medicine, perhaps we need to think beyond biological interventions. Like the moon, we seem a long way away from provable therapies to increase human healthspan (the healthy period of one's life) or lifespan (how long one lives). This review therefore focuses on radical proposals. We hope it might stimulate discussion on what we might consider doing significantly differently than ongoing aging research.

**Introduction – incrementalism & thinking linear vs. in log-scale**
Silicon Valley wants 10X engineers, that is people who do 10X the work of the average engineer. Likewise, biomedical researchers starting their careers are told by their senior colleagues to chase big findings and ignore small effects. To aim for big increases in human healthspan or lifespan (assuming the Earth can handle the consequences and enough people want to), let's first consider the current state of aging delaying therapies.

There are many ways to think about treating aging. Most existing treatments treat specific aging conditions. They are in the form of drugs that improve one area of the body that is diseased in aging populations. Drugs for diabetes (high blood glucose) and osteoporosis (bone loss) are common examples. Currently, there are no FDA-approved remedies to reduce aging per se, i.e., to increase healthspan or lifespan. Experiments aimed at identifying new interventions for aging have historically been done in species ranging from primates to mice to worms to yeast (1). Yet, the pharmaceutical industry has become weary of treatments that work in model systems only to be proven to be ineffective in humans (2). This is unfortunately especially true for treatments for aging-associated chronic diseases (3). Thus, considering human clinical trials for aging therapies will take years to play out, it is important to scrutinize the effect sizes seen in preclinical models.

A less than encouraging sign for many of the lifespan experiments done in mammals is that they have modest effect sizes, often only having statistically significant effects in one of the genders, and often only in specific dietary and/or housing conditions (4). Even inhibiting one of the most potent aging pathways (5-7), the mechanistic target of rapamycin (mTOR) pathway (8) has arguably modest effects on lifespan – a 12-24% increase in mice (14-23% in females/9-26% in males) (9, 10). Notably, rapamycin has been associated with harmful effects, including increased risk of aggressive hematological malignancies and dramatic microbiome alterations (11). Also, these experiments were done in a controlled environment which limits pathogens exposure. This is relevant because rapamycin is an immunosuppressant, used clinically in organ transplantation (12). This suggests in a real-world scenario where people's immune systems would be challenged regularly, it's not clear rapamycin would have a significant beneficial effect on healthspan or lifespan.

This is all to ask, if rapamycin is one of the potential best-case scenarios and might be predicted to have a modest effect if any (and possibly a detrimental one) in people, should it continue to receive so much focus by the aging community? Note the problems in the aging field with small and inconsistent effects for the leading strategies aren't specific to rapamycin. The popular reactive oxygen species (ROS) theory of aging (13) was challenged when it was found that deleting the genes capable of detoxifying superoxide are dispensable for worm lifespan (14). In a related example in worms, different labs produced dramatically different lifespan-modulating effect sizes with the same with ROS-altering drug (15). That worms are simpler than mice and it took over at decade by highly qualified labs to resolve the inconsistent results with the ROS drug, again doesn't bode well for ROS-based therapies in the real world with people.

Treating individual aging-related diseases has encountered roadblocks that should also call into question whether we are on the optimal path for human aging. Alzheimer's and diabetes are particularly well-funded and well-researched aging-related topics where there are still huge gaps in our understanding and lack of good treatment options. With Alzheimer's there has been considerable focus on amyloid beta and tau, but targeting those molecules hasn't done much for Alzheimer's so far leaving many searching for answers (16, 17). Regarding diabetes, some people think insulin resistance is caused by intracellular over-accumulation of lipids, but despite many smart people working on it for decades, it isn't clear whether diacylglycerol (18), ceramide (19), or other lipids might be the main culprit. The point is when we spend collectively a long time on something that isn't working well, such as manipulating a single gene or biological process, it should seem natural to consider conceptually different approaches. Herein we will discuss different therapeutic concepts for aging that might push us beyond our current thinking.

**Beyond the reductionist one-hypothesis, one-pathway, one-disease view.**
Perhaps the problem is that aging might not be a single molecular pathway or collection of pathways as we biologists typically understand pathways. Following the omnigenic model of complex disease (20), aging might involve all genes. Therefore, if this is true, would doing one thing to extend life, such as inhibiting one pathway with a drug, really be expected to do a lot?

Still, part of the omnigenic model is that there are "hub"/"core" genes that are relatively more important than other genes. It is intuitive to target the genes we consider to be most impactful, so if one drug or manipulating one gene may not be good enough, looking into synergistic combinations as we do with treating AIDS (21) or cancer (22, 23) seems like a next logical step.

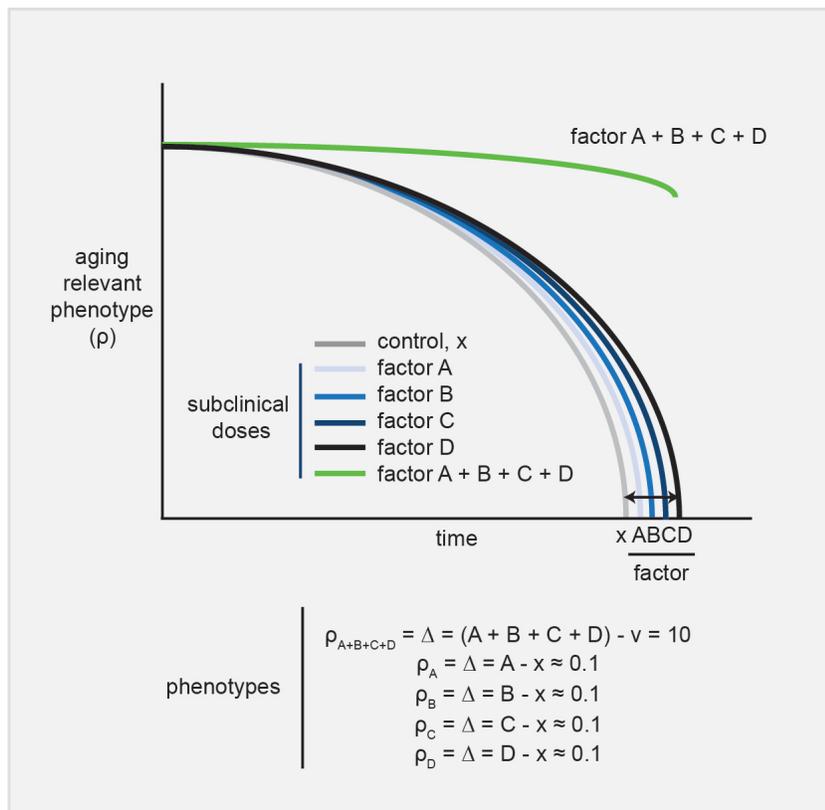

**Figure 1: 10X healthspan/lifespan strategy #1. Identify maximally synergistic life-extending factors, such as drugs or gene manipulations.** Four factors are depicted as needed to promote synergistic effects on an aging phenotype, ρ (rho), in analogy to the Yamuna pluripotency cocktail, which contains four factors. These synergistic factors, which could be drugs and/or gene manipulations, will ideally be chosen from a large list of candidates following systematic, unbiased, large-scale experimentation.

One proof-of-concept, aging-limiting cocktail is the Yamanaka pluripotency factors (24) that was tested with good results in a mouse model of the premature aging condition, progeria (25). Manipulating both insulin-like growth factor (IGF) and sterol-responsive element binding protein (SREBP) also produced strong synergistic healthspan-extending effects (26). Several other combinations have likewise been validated in model organisms (27, 28).

Using synergistic combinations has the potential to lower the amount of each agent one has to use, which should be beneficial in lowering their potential side effects (29, 30). Still, the effect sizes from combinations of a few factors might not be much more than those of a single factor. To even begin to contemplate 10X effects like Silicon Valley aims for, we need to concentrate on the strategies that will generate the most synergistic combinations (Figure 1). This is a challenge because of the number of possible combinations to be tested is a multiple of the number of factors under consideration. Consider the hypothetical that we had drugs targeting each of the ~20,000 protein-coding human genes. Testing all those in combination would be four hundred million combinations, which isn't feasible today (noting that in some contexts the effects of combinations of many factors can be approximated by the effects of pairs, which reduces the number of combinations to test (31)). Potentially we might not need to choose and we could just throw the kitchen sink at the problem and perturb many genes at a time. There is precedence for the many-gene approach using CRISPR, where 62 retroviral genes were knocked out in the pig liver to make these organs theoretically safer for transplantation (32). Yet there are regulatory hurdles that need to be cleared. Currently the FDA requires testing of the individual components of a combination treatment. However, one can envision a future where the need will get so great that the FDA will allow applications for cocktails without their individual components gaining approval first (33).

Even thinking of specific aging-associated conditions rather than aging per se, synergistic factors could provide a valuable, new path forward. Because the same core genes tend to be important across aging conditions (34) this suggests there would be a benefit for researchers interested in specific conditions to look at aging per se or other aging conditions besides the one they're studying to better understand them. For physicians, it might also mean when the one-drug approach is failing it like it is with Alzheimer's, they might consider drug synergies with drugs that alleviate other aging-associated conditions. This would be a form of drug repositioning (35) that already has evidence it could work for Alzheimer's. For example, drugs for several aging-related conditions such as heart disease (36, 37), diabetes (38), osteoporosis (39) have all been shown to alleviate neurodegeneration in various models.

Yamanaka started his pursuit of a pluripotency cocktail with a short list of genes. If we were to make a short list of molecular factors for aging what would it look like? On this point, we'd like to make a plug for factors that add back function to aging tissues. Many existing strategies to increase healthspan or lifespan involve taking something away, often by inhibiting molecular pathways. Dietary limitation such as caloric restriction and intermittent fasting are the most-well known approaches in this category (40, 41). Both are dietary regimens that involve reducing calorie intake without causing malnutrition. Though caloric restriction is a long-standing "gold standard" in the aging field, and is associated with up to 50% increases in lifespan (1), it has two major issues: 1) It has poor feasibility. It is questionable whether many humans could cut the large number of calories, e.g., 33% reduction, needed long-term to demonstrate the generalizable efficacy of this approach (42); 2) more importantly, though many studies including those dating back almost a century ago have showed significant benefits of CR (43-45), many other studies haven't panned out (46-48). How can CR work so well in some contexts and not at all in others? Again, like rapamycin, one has to wonder whether rather than devote more resources to CR, there might be other better approaches.

Therefore, instead of restricting something, arguably a more defensible idea would be to restore functionality. This makes sense considering aging explicitly means decreasing function. How would one simultaneously restore declining molecular functioning of various tissues, such as the bone (49), immune system (50), and heart (51)? Parabiosis involving the sharing of blood streams between young and old mice provides a proof-of-concept of for how we might add back combinations of factors to restore functionality throughout the body (52). Though having a renewable, multiplexed source of "young" signaling that doesn't require another human's involvement, such as using extracellular vesicles (53), would seem a more feasible approach.

**Is measuring lifespan the right experimental model?**
As is the case in many areas of molecular biology, some models get a lot of attention and others get less. Measuring organismal lifespan – the time from birth to death – is a core assay in the aging field. Why then are we not studying trees or other examples of extreme longevity more (54)? Trees are the longest-lived species on earth living up to 5,000 years. They have stem cells that actively divide and differentiate throughout their existence (55). They have many of the same hub/core genes that humans have (56). They are enriched for pathways elevated in young mammalian tissues relative to their older counterparts (57, 58). Therefore, shouldn't there be principles there we could learn more from? The mainstreaming of the long-lived naked mole rat and various aquatic species over the last few decades has brought many new ideas to the aging field (59, 60). Though some are starting to the study longevity determinants in trees (61), it is safe to say they are underexplored.

Measuring lifespan makes sense because we are measuring the thing we want to improve. However, can we identify factors that might extend life without measuring lifespan? In worms, being alive means the worm moves, even if only slightly. Similarly, in humans, a key symbolic if not medical part of being considered alive means being able to respond to external cues. Both worms and people being considered alive says nothing about quality of life though. In worms, yes, one can disrupt mitochondria function and that can extend lifespan in model organisms significantly, but would inhibiting such a key organelle long-term in humans resemble anything like a normal life?

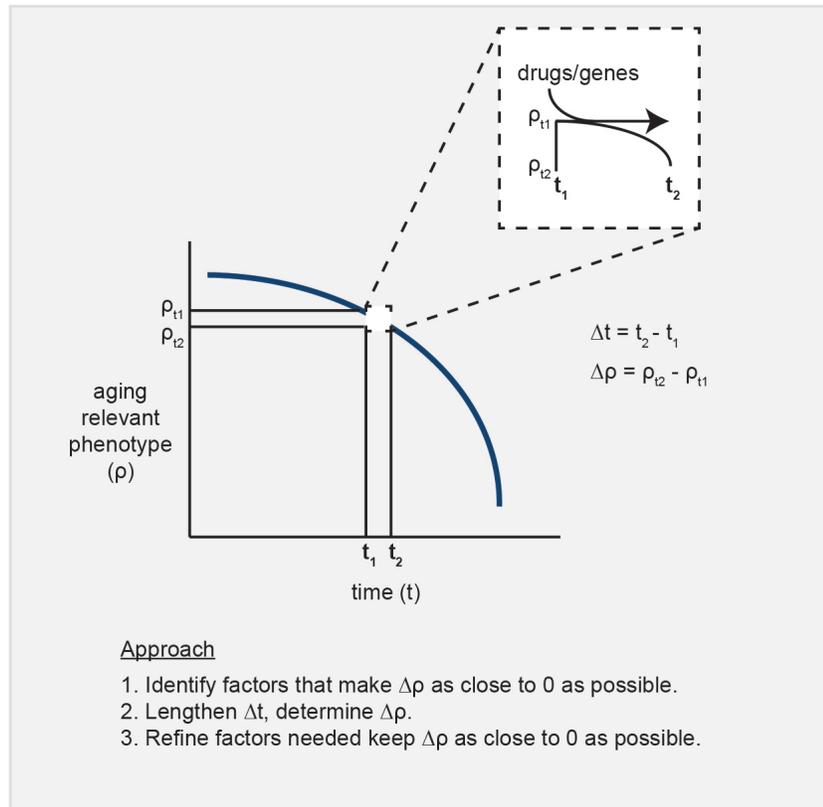

Figure 2: 10X healthspan/lifespan strategy #2. Minimize aging-related changes on a small timescale. Rather than waiting a lifespan to observe the effects of a putative aging-delaying/healthspan-increasing factor(s), one can measure an aging-relevant phenotype, $\rho$, on small time scale (e.g., hours) and identify changes that make the rate of change, i.e., derivative ($\Delta\rho/\Delta t$), go from negative to 0. This requires highly quantitative, rigorous experimentation.

In the spirit of pioneers, explorers, and astronauts who couldn't know where they were headed, instead of measuring lifespan, perhaps one could aim for simpler goals. For example, we could figure out how to maintain a young state on an incremental timescale. The younger vs. older states could be defined, for example, by determining the mRNA levels of various genes at each time point. The key would be being able to detect slight differences in the levels of the factors of interest (i.e., biomarkers) between the two time points (Figure 2). Then once the young phenotype is preserved on a reasonable timescale, we would seek to extrapolate to longer time scales. This is, in essence, a healthspan experiment rather than a lifespan experiment. Thus, if aren't measuring lifespan, what aging-relevant phenotypes should we look at? Replicative life span, which is the number of divisions a cell will undergo before it dies, decreases with chronological age of the organism (62). Therefore, measuring cell replication provides a simple model to evaluate small differences in the young vs. slightly older state. Cell number is an experimentally tractable phenotype, and inexpensive to measure too. Though cell replication rates are only one aspect of age and as many in the aging field know this might be considered a solved issue as telomerase can confer immortal growth (63). However, telomerase is re-activated in 90% of cancers and influences DNA damage responses (64) but doesn't influence all core/hub genes. Thus, to identify robust, aging-relevant phenotypes, we seem due for a comprehensive assessment of the molecular differences between cells/tissues as they age.

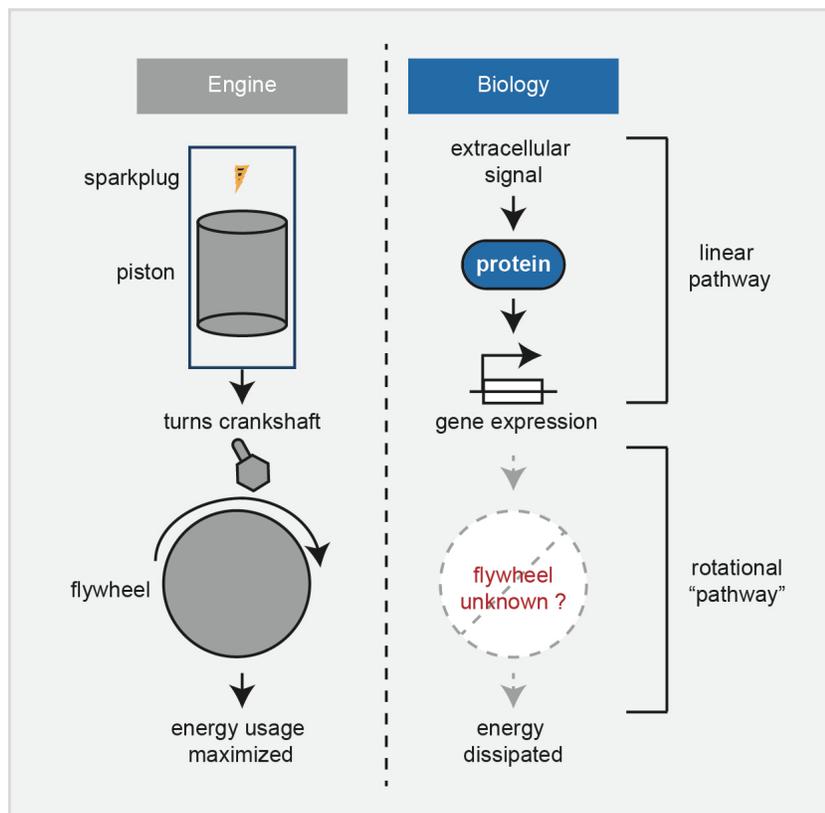

**Figure 3: 10X healthspan/lifespan strategy #3. Identify flywheels that preserve our body's energy.** Engines work by gas injected into its cylinders on regular intervals being ignited by a sparkplug spark. The gas is volatilized and expands, driving the piston, which then turns the crankshaft. The crankshaft is attached to a flywheel and the wheels. The flywheel has a large moment of inertia which maintains its rotation in between the time interval where the sparkplug fires. Therefore, the flywheel both maximizes the energy derived from the gas and smoothens the ride. It is currently unclear whether any biomolecules or non-molecular factors function as a flywheel for humans.

**Beyond molecular biology**
Often discoveries in one field are made using findings from orthogonal fields. The contemporary internet stemming from efforts by the military to build a communication network is a classic example (65). What principles from other disciplines could be employed to help biologist tackle aging? Around the time of World War II, physicists flooded into biology to help solve problems such as the structure of DNA. Perhaps we could use their help again with aging.

Thinking as an engineer, consider our bodies as a car engine. While most of the aging biologist's focus has been on combustion and its linear pathway of energy generation, less attention has been on the flywheel (66), which maximizes the energy from combustion via its rotational momentum (Figure 3). Flywheels also ensure a smoother ride such that when the sparkplug sparks, we don't feel a jolt due to the discontinuous action of the piston. This engine analogy seems apt to aging because many in the aging field have focused on the cellular sparkplugs, such as the mitochondria, with less effort on what could be our flywheel. Moreover, molecular biology is only one way to make a flywheel. Perhaps we can generalize the inputs to our body's engine to include more factors than just biomolecules. Car flywheel moment of inertias have been optimized to maximize fuel efficiency. What besides molecules might diminish our moment of inertia as we get older?

Certainly, physical chemistry is underexplored for its role in aging. Could we make any environmental changes (not detrimental to our planet) that could keep our flywheels spinning? Is our body temperature and the atmosphere we experience optimized for our healthspans and lifespans? The effects of varying atmospheric gases has been tested in the context of endurance athletes (67) and with divers (68, 69), but to our knowledge not in the context of aging. We know molecular oxygen is important but what about other components of our atmosphere? It is surprising how few studies have measured the effects of the most abundant atmospheric gas, nitrogen, on human physiology (70). Several species such as *Geogemma barossii* live at extreme temperatures (250 degrees) (71). There are some correlations between external and internal temperatures and longevity in humans, but this should be tested more systematically.

What about the potential, electrical, and mechanical energy of aging? We know lack of gravity leads to decline in our muscles and bone that resembles aging and mechanical forces such as exercise help delay aging (72, 73). Though exercising our muscles and bones seems an inadequate way to fix all our aging tissues. Is there a way to deliver mechanical forces to other organs beyond our muscle and bones? Can we exercise the liver? We have to be careful with the type of mechanical forces though because in tissues such as the liver, blood flow congestion induces fibrosis that impairs its function (74).

There are also many man-made factors that profoundly affect aging. Social isolation strongly negatively affects our health (75). Socioeconomic factors are too broad a discussion for this forum, but it shouldn't be a stretch to say focusing on them might possibly produce larger global health benefits than many medical therapies.

**Looking ahead**
In summary, we laid out the first steps for three potential strategies that we believe, if optimized, might have the potential to increase human healthspan and/or lifespan significantly beyond what's currently possible: 1) identify factors that strongly synergize with each other, ideally without regard to number of factors that need to be combined; 2) identify factors that prevent age-associated changes over tractable time increments and that retain their prophylactic effects as the time interval is lengthened; 3) identify factors that play a flywheel function for our bodies to make them more energy efficient and balance fluctuations in our physiology. This review is a thought exercise on what could we do if we could study anything we think might help with aging. At the same time, constraints can also drive progress. In that light, what would we do to combat aging if our knowledge of molecular biology didn't exist?

# REFERENCES

1. Fontana L, Partridge L, Longo VD. Extending healthy life span--from yeast to humans. Science. 2010;328(5976):321-6.
2. Scannell JW, Blanckley A, Boldon H, Warrington B. Diagnosing the decline in pharmaceutical R&D efficiency. Nat Rev Drug Discov. 2012;11(3):191-200.
3. Barzilai N, Cuervo AM, Austad S. Aging as a Biological Target for Prevention and Therapy. JAMA. 2018;320(13):1321-2.
4. Austad SN. The geroscience hypothesis: is it possible to change the rate of aging? In: Sierra F, Kohanski R, eds.2015.
5. Lopez-Otin C, Blasco MA, Partridge L, Serrano M, Kroemer G. The hallmarks of aging. Cell. 2013;153(6):1194-217.
6. Kennedy BK, Berger SL, Brunet A, Campisi J, Cuervo AM, Epel ES, et al. Geroscience: linking aging to chronic disease. Cell. 2014;159(4):709-13.
7. Sierra F, Kohanski, R. Advances in Geroscience2016.
8. Kennedy BK, Lamming DW. The Mechanistic Target of Rapamycin: The Grand ConducTOR of Metabolism and Aging. Cell Metab. 2016;23(6):990-1003.
9. Harrison DE, Strong R, Sharp ZD, Nelson JF, Astle CM, Flurkey K, et al. Rapamycin fed late in life extends lifespan in genetically heterogeneous mice. Nature. 2009;460(7253):392-5.
10. Miller RA, Harrison DE, Astle CM, Fernandez E, Flurkey K, Han M, et al. Rapamycin-mediated lifespan increase in mice is dose and sex dependent and metabolically distinct from dietary restriction. Aging Cell. 2014;13(3):468-77.
11. Bitto A, Ito TK, Pineda VV, LeTexier NJ, Huang HZ, Sutlief E, et al. Transient rapamycin treatment can increase lifespan and healthspan in middle-aged mice. Elife. 2016;5.
12. Waldner M, Fantus D, Solari M, Thomson AW. New perspectives on mTOR inhibitors (rapamycin, rapalogs and TORKinibs) in transplantation. Br J Clin Pharmacol. 2016;82(5):1158-70.
13. Harman D. Aging: a theory based on free radical and radiation chemistry. J Gerontol. 1956;11(3):298-300.
14. Van Raamsdonk JM, Hekimi S. Superoxide dismutase is dispensable for normal animal lifespan. Proc Natl Acad Sci U S A. 2012;109(15):5785-90.
15. Lithgow GJ, Driscoll M, Phillips P. A long journey to reproducible results. Nature. 2017;548(7668):387-8.
16. Mehta D, Jackson R, Paul G, Shi J, Sabbagh M. Why do trials for Alzheimer's disease drugs keep failing? A discontinued drug perspective for 2010-2015. Expert Opin Investig Drugs. 2017;26(6):735-9.
17. Murphy MP. Amyloid-Beta Solubility in the Treatment of Alzheimer's Disease. N Engl J Med. 2018;378(4):391-2.
18. Erion DM, Shulman GI. Diacylglycerol-mediated insulin resistance. Nat Med. 2010;16(4):400-2.
19. Meikle PJ, Summers SA. Sphingolipids and phospholipids in insulin resistance and related metabolic disorders. Nat Rev Endocrinol. 2017;13(2):79-91.
20. Boyle EA, Li YI, Pritchard JK. An Expanded View of Complex Traits: From Polygenic to Omnigenic. Cell. 2017;169(7):1177-86.
21. Larder BA, Kemp SD, Harrigan PR. Potential mechanism for sustained antiretroviral efficacy of AZT-3TC combination therapy. Science. 1995;269(5224):696-9.
22. O'Neil NJ, Bailey ML, Hieter P. Synthetic lethality and cancer. Nat Rev Genet. 2017;18(10):613-23.
23. Humphrey RW, Brockway-Lunardi LM, Bonk DT, Dohoney KM, Doroshow JH, Meech SJ, et al. Opportunities and challenges in the development of experimental drug combinations for cancer. J Natl Cancer Inst. 2011;103(16):1222-6.
24. Takahashi K, Yamanaka S. Induction of pluripotent stem cells from mouse embryonic and adult fibroblast cultures by defined factors. Cell. 2006;126(4):663-76.
25. Ocampo A, Reddy P, Martinez-Redondo P, Platero-Luengo A, Hatanaka F, Hishida T, et al. In Vivo Amelioration of Age-Associated Hallmarks by Partial Reprogramming. Cell. 2016;167(7):1719-33 e12.
26. Admasu TD, Chaithanya Batchu K, Barardo D, Ng LF, Lam VYM, Xiao L, et al. Drug Synergy Slows Aging and Improves Healthspan through IGF and SREBP Lipid Signaling. Dev Cell. 2018;47(1):67-79 e5.
27. Evason K, Huang C, Yamben I, Covey DF, Kornfeld K. Anticonvulsant medications extend worm life-span. Science. 2005;307(5707):258-62.

28. Chen D, Li PW, Goldstein BA, Cai W, Thomas EL, Chen F, et al. Germline signaling mediates the synergistically prolonged longevity produced by double mutations in daf-2 and rsks-1 in C. elegans. Cell Rep. 2013;5(6):1600-10.
29. Foucquier J, Guedj M. Analysis of drug combinations: current methodological landscape. Pharmacol Res Perspect. 2015;3(3):e00149.
30. Keith CT, Borisy AA, Stockwell BR. Multicomponent therapeutics for networked systems. Nat Rev Drug Discov. 2005;4(1):71-8.
31. Wood KB. Pairwise interactions and the battle against combinatorics in multidrug therapies. Proc Natl Acad Sci U S A. 2016;113(37):10231-3.
32. Yang L, Guell M, Niu D, George H, Lesha E, Grishin D, et al. Genome-wide inactivation of porcine endogenous retroviruses (PERVs). Science. 2015;350(6264):1101-4.
33. Choi SH, Wang Y, Conti DS, Raney SG, Delvadia R, Leboeuf AA, et al. Generic drug device combination products: Regulatory and scientific considerations. Int J Pharm. 2018;544(2):443-54.
34. Kammenga JE. The background puzzle: how identical mutations in the same gene lead to different disease symptoms. FEBS J. 2017;284(20):3362-73.
35. Pushpakom S, Iorio F, Eyers PA, Escott KJ, Hopper S, Wells A, et al. Drug repurposing: progress, challenges and recommendations. Nat Rev Drug Discov. 2018.
36. Williams PT. Lower risk of Alzheimer's disease mortality with exercise, statin, and fruit intake. J Alzheimers Dis. 2015;44(4):1121-9.
37. Daneschvar HL, Aronson MD, Smetana GW. Do statins prevent Alzheimer's disease? A narrative review. Eur J Intern Med. 2015;26(9):666-9.
38. Markowicz-Piasecka M, Sikora J, Szydlowska A, Skupien A, Mikiciuk-Olasik E, Huttunen KM. Metformin - a Future Therapy for Neurodegenerative Diseases : Theme: Drug Discovery, Development and Delivery in Alzheimer's Disease Guest Editor: Davide Brambilla. Pharm Res. 2017;34(12):2614-27.
39. Tiihonen M, Taipale H, Tanskanen A, Tiihonen J, Hartikainen S. Incidence and Duration of Cumulative Bisphosphonate Use among Community-Dwelling Persons with or without Alzheimer's Disease. J Alzheimers Dis. 2016;52(1):127-32.
40. Most J, Tosti V, Redman LM, Fontana L. Calorie restriction in humans: An update. Ageing Res Rev. 2017;39:36-45.
41. Longo VD, Panda S. Fasting, Circadian Rhythms, and Time-Restricted Feeding in Healthy Lifespan. Cell Metab. 2016;23(6):1048-59.
42. Minor RK, Allard JS, Younts CM, Ward TM, de Cabo R. Dietary interventions to extend life span and health span based on calorie restriction. J Gerontol A Biol Sci Med Sci. 2010;65(7):695-703.
43. McCay CM, Crowell MF, Maynard LA. The effect of retarded growth upon the length of life span and upon the ultimate body size. 1935. Nutrition. 1989;5(3):155-71; discussion 72.
44. Colman RJ, Anderson RM, Johnson SC, Kastman EK, Kosmatka KJ, Beasley TM, et al. Caloric restriction delays disease onset and mortality in rhesus monkeys. Science. 2009;325(5937):201-4.
45. Colman RJ, Beasley TM, Kemnitz JW, Johnson SC, Weindruch R, Anderson RM. Caloric restriction reduces age-related and all-cause mortality in rhesus monkeys. Nat Commun. 2014;5:3557.
46. Vaughan KL, Kaiser T, Peaden R, Anson RM, de Cabo R, Mattison JA. Caloric Restriction Study Design Limitations in Rodent and Nonhuman Primate Studies. J Gerontol A Biol Sci Med Sci. 2017;73(1):48-53.
47. Mattison JA, Roth GS, Beasley TM, Tilmont EM, Handy AM, Herbert RL, et al. Impact of caloric restriction on health and survival in rhesus monkeys from the NIA study. Nature. 2012;489(7415):318-21.
48. Harper JM, Leathers CW, Austad SN. Does caloric restriction extend life in wild mice? Aging Cell. 2006;5(6):441-9.
49. Ribault D, Habib M, Abdel-Majid K, Barbara A, Mitrovic D. Age-related decrease in the responsiveness of rat articular chondrocytes to EGF is associated with diminished number and affinity for the ligand of cell surface binding sites. Mech Ageing Dev. 1998;100(1):25-40.
50. de Haan G, Van Zant G. Dynamic changes in mouse hematopoietic stem cell numbers during aging. Blood. 1999;93(10):3294-301.
51. Gerstenblith G, Spurgeon HA, Froehlich JP, Weisfeldt ML, Lakatta EG. Diminished inotropic responsiveness to ouabain in aged rat myocardium. Circ Res. 1979;44(4):517-23.
52. Villeda SA, Luo J, Mosher KI, Zou B, Britschgi M, Bieri G, et al. The ageing systemic milieu negatively regulates neurogenesis and cognitive function. Nature. 2011;477(7362):90-4.
53. Robbins PD. Extracellular vesicles and aging. Stem Cell Investig. 2017;4:98.

54. Anson RM, Willcox B, Austad S, Perls T. Within- and between-species study of extreme longevity--comments, commonalities, and goals. J Gerontol A Biol Sci Med Sci. 2012;67(4):347-50.
55. Flanary BE, Kletetschka G. Analysis of telomere length and telomerase activity in tree species of various lifespans, and with age in the bristlecone pine Pinus longaeva. Rejuvenation Res. 2006;9(1):61-3.
56. Altenhoff AM, Glover NM, Train CM, Kaleb K, Warwick Vesztrocy A, Dylus D, et al. The OMA orthology database in 2018: retrieving evolutionary relationships among all domains of life through richer web and programmatic interfaces. Nucleic Acids Res. 2018;46(D1):D477-D85.
57. Peterson TR. Gene expression patterns in young versus old tissues. 2018.
58. Xu YH, Liao YC, Zhang Z, Liu J, Sun PW, Gao ZH, et al. Jasmonic acid is a crucial signal transducer in heat shock induced sesquiterpene formation in Aquilaria sinensis. Sci Rep. 2016;6:21843.
59. Lagunas-Rangel FA, Chavez-Valencia V. Learning of nature: The curious case of the naked mole rat. Mech Ageing Dev. 2017;164:76-81.
60. Austad SN. Methusaleh's Zoo: how nature provides us with clues for extending human health span. J Comp Pathol. 2010;142 Suppl 1:S10-21.
61. Plomion C, Aury JM, Amselem J, Leroy T, Murat F, Duplessis S, et al. Oak genome reveals facets of long lifespan. Nat Plants. 2018;4(7):440-52.
62. Lee JS, Lee SM, Jeong SW, Sung YG, Lee JH, Kim KW. Effects of age, replicative lifespan and growth rate of human nucleus pulposus cells on selecting age range for cell-based biological therapies for degenerative disc diseases. Biotech Histochem. 2016;91(5):377-85.
63. Blackburn EH, Epel ES, Lin J. Human telomere biology: A contributory and interactive factor in aging, disease risks, and protection. Science. 2015;350(6265):1193-8.
64. Masutomi K, Possemato R, Wong JM, Currier JL, Tothova Z, Manola JB, et al. The telomerase reverse transcriptase regulates chromatin state and DNA damage responses. Proc Natl Acad Sci U S A. 2005;102(23):8222-7.
65. https://en.wikipedia.org/wiki/History_of_the_Internet.
66. https://en.wikipedia.org/wiki/Flywheel.
67. Levine BD, Stray-Gundersen J. Dose-response of altitude training: how much altitude is enough? Adv Exp Med Biol. 2006;588:233-47.
68. Freiberger JJ, Derrick BJ, Natoli MJ, Akushevich I, Schinazi EA, Parker C, et al. Assessment of the interaction of hyperbaric N2, CO2, and O2 on psychomotor performance in divers. J Appl Physiol (1985). 2016;121(4):953-64.
69. Bove AA. Diving medicine. Am J Respir Crit Care Med. 2014;189(12):1479-86.
70. Harmens H, Hayes F, Sharps K, Mills G, Calatayud V. Leaf traits and photosynthetic responses of Betula pendula saplings to a range of ground-level ozone concentrations at a range of nitrogen loads. J Plant Physiol. 2017;211:42-52.
71. Kashefi K, Lovley DR. Extending the upper temperature limit for life. Science. 2003;301(5635):934.
72. Turner CH, Takano Y, Owan I. Aging changes mechanical loading thresholds for bone formation in rats. J Bone Miner Res. 1995;10(10):1544-9.
73. Garg K, Boppart MD. Influence of exercise and aging on extracellular matrix composition in the skeletal muscle stem cell niche. J Appl Physiol (1985). 2016;121(5):1053-8.
74. Simonetto DA, Yang HY, Yin M, de Assuncao TM, Kwon JH, Hilscher M, et al. Chronic passive venous congestion drives hepatic fibrogenesis via sinusoidal thrombosis and mechanical forces. Hepatology. 2015;61(2):648-59.
75. Valtorta N, Hanratty B. Loneliness, isolation and the health of older adults: do we need a new research agenda? J R Soc Med. 2012;105(12):518-22.